\documentclass[aps,prb,twocolumn,floats]{revtex4-1}
\usepackage{graphicx}
\usepackage{amsmath}
\usepackage{amssymb}
\usepackage{bm}
\usepackage{mathtools}
\usepackage[normalem]{ulem}
\usepackage{color}
\usepackage{tikz}
\usepackage{textcomp}
\usepackage[colorlinks,linkcolor=red,citecolor=blue,filecolor=magenta,
urlcolor=cyan]{hyperref}

\begin{document}
\title{Intrinsic and extrinsic geometries of correlated many-body states}
\author{Ankita Chakrabarti}
\email{ankitac@imsc.res.in}
\affiliation{The Institute of Mathematical Sciences, C.I.T. Campus, Chennai 600 
113, India}
\affiliation{Homi Bhabha National Institute, Training School Complex, Anushakti Nagar, Mumbai 400094, India}
\author{S. R. Hassan}
\email{shassan@imsc.res.in}
\affiliation{The Institute of Mathematical Sciences, C.I.T. Campus, Chennai 600 
113, India}
\affiliation{Homi Bhabha National Institute, Training School Complex, Anushakti Nagar, Mumbai 400094, India}
\author{R. Shankar}
\email{shankar@imsc.res.in}
\affiliation{The Institute of Mathematical Sciences, C.I.T. Campus, Chennai 600 
113, India}
\affiliation{Homi Bhabha National Institute, Training School Complex, Anushakti Nagar, Mumbai 400094, India}

\date{\today}

\begin{abstract}
We explore two approaches to characterise the quantum geometry of the ground
state of correlated fermions in terms of the distance matrix in the spectral
parameter space. (a) An intrinsic geometry approach, in which we study the
intrinsic curvature defined in terms of the distance matrix. (b) An extrinsic
geometry approach, in which we investigate how the distance matrix can be
approximately embedded in finite dimensional Euclidean spaces. We implement
these approaches for the ground state of a system of one-dimensional fermions
on a 18-site lattice with nearest neighbour repulsion. The intrinsic curvature
sharply changes around the Fermi points in the metallic regime but is more or
less uniform in the insulating regime. In the metallic regime, the embedded
points clump into two well seperated sets, one corresponding to modes in the
Fermi sea and the other to the modes outside it. In the insulating regime, the
two sets tend to merge.
\end{abstract}
\maketitle
\section{Introduction}

In the past few decades, there has been a body of work attempting to 
characterise many-body states using the concepts of quantum geometry 
\cite{RestaSorella,Resta1,Haldane2004}. The symmetries of the ground states of 
many-body systems, characterised by the expectation values of local
order parameters have proven to be a very useful way to characterise
the phases of the systems \cite{Landau:1937obd}. However, in several
cases this is inadequate. There can be two phases of the system with very
different physical properties but with the same symmetry. The metallic
and insulating phases of a solid are the prime example of such a situation. 
These two phases can be distinguished by the nature of the excited states
of the system. The metallic state has gapless charged excitations whereas the 
charged excitations in the insulating phase have a gap. 

In a seminal paper, Walter Kohn \cite{Walterkohn} suggested that it is
possible to distinguish these two phases by examining the structure of the
ground state alone. Namely, given only the ground state, it is possible to
predict if the system is metallic or insulating. Kohn also gave a qualitative 
description of what distinguishes the ground state of a metal and an insulator.
This idea has been followed up by others \cite{L1,RestaSorella,Resta1,Resta2011}, and 
it has been been argued that this difference can be quantified using the 
concepts of quantum geometry. 

In particular, it has been shown that the so called localisation tensor, the
second moment of the pair correlation function in the ground state is a
quantity that distinguishes between the two phases and characterises the
insulating phase. It is divergent in the metallic phase and finite in the
insulating phase. Further its value in the insulating phase is a measure of the
localisation of the electrons \cite{RestaSorella,SouzaMartin,Resta2011}. It has also been shown 
that for mean field states corresponding to filled bands, the localisation tensor is
the integral of the quantum metric corresponding to the filled bands over the 
Brilliuon zone \cite{Resta1,SCPMR}. This result establishes the relation of the 
metallic/insulating property of the system to the quantum geometry for 
band insulators which are well described by mean field states.

In previous work \cite{Paper1}, we had given a definition of quantum
distances in the space of spectral parameters for general many particle states.
For fermions in a periodic potential, the spectral parameters are the
quasi-momenta in the Brillioun zone.  We had shown that our definition
satisfies the triangle inequalities and reduces to the standard definition, in
terms of the single particle wave functions for mean-field states.

An important feature of our definition is that it gives sensible and nontrivial
results for the distance matrix in the space of spectral parameters for systems
with partially filled bands. Thus, unlike the previous studies, our formalism
enables us to probe the quantum geometry of many-body states both in the
metallic phase and in the insulating phase.

In this paper, we attempt to find ways to characterise the quantum geometry
of many-fermion states in terms of its distance matrix. We explore two
approaches. First, we study the intrinsic curvature implied by the distance
matrix.  Next, we study an approximate embedding of the distance matrix in low
dimensional Euclidean spaces which gives a visualization of the ground state. 

How are these geometric quantities related to the physics of the system? We do
not have general results on this issue just now, but discuss it in the context
of the known physics of a specific, well studied, model of interacting
fermions, the one-dimensional $t$-$V$ model at half filling, a model of spinless
fermions on a lattice with nearest neighbour repulsion \cite{Yang-XXZ,Baxter,tV-bosonisation}. 
At $V/t=2$, this model has a metal-insulator transition \cite{Shankar}. At $V=0$ it is 
a trivial Fermi liquid, at $0<V/t<2$ it is a metallic Luttinger liquid and at $V/t>2$, 
it is an insulator. We solve this model numerically for 18 sites. Since it is a finite
size system, there is no phase transition but only a crossover. We compute  the
different geometric quantities that characterise the ground state and analyse
how they differ in the two regimes. 

The rest of the paper is organised as follows. In Section \ref{convention}, we
review our definition of quantum distances in the spectral parameter space. In
the model we study, these parameters are the quasi-momenta taking values in the
Brillioun zone ($BZ$). As mentioned above, we are studying finite size model.
Thus the quasi-momenta are discrete and finite. We therefore need the
techniques of discrete geometry to study the system. The model and the
mathematical notations to study its geometry are also described in this
section.  The definition that we use for the intrinsic curvature of a discrete
set is discussed in Section \ref{curvature}.  This section also presents the
results of the computation of the curvature in our model and its relation to
the physics of the system.  In Section \ref{ee} we motivate our attempts to
approximately embed the distance matrix in a Euclidean space. In Section \ref{mfsee},
we review the fact that, for mean field states, our definition of quantum distances
reduces to the standard Hilbert-Schmidt distances in terms of the single
particle wave functions.  We then show that, in general, Hilbert-Schmidt
distances in a finite dimensional Hilbert space can be isometrically embedded
in a finite dimensional Euclidean space. In Section \ref{genee}, we review the techniques of
embedding a general distance matrix in a Euclidean space. Section
\ref{edmextremes} implements the procedure analytically for the two extreme
limits of the coupling constant, $V=0,\infty$.  The numerical results for
finite, non-zero $V$ are presented in Section \ref{edmfinite}. Section \ref{approxD} 
discusses the concept of approximate embedding in finite dimensional
Euclidean spaces. Section \ref{edmw} discusses the Euclidean embedding for the
so called Wasserstein distance matrix, a quantity we use to define the
intrinsic curvature. We summarise our results and discuss the conclusions we
draw from them in Section \ref{concl}.

\section{Definitions, conventions and notation}
\label{convention}

In this section we first review our definition of the distance matrix between
pair of points in the spectral parameter space  for correlated states \cite{Paper1}. 
We define the notion of a graph associated with a state of a finite dimensional system. 
We then describe the details of the model that we study by numerical exact diagonalization.

\subsection{The distance matrix}
\label{DMintro}

In previous work \cite{Paper1}, we had defined the quantum distance $d(\mathbf k_i,\mathbf
k_j)$  between two points  $\mathbf k_i$ and $\mathbf k_j$ in the spectral
parameter space in terms of the expectation values of what we called the
exchange operators. We review that definition below.

We consider a tight binding model on a Bravais lattice with $N_B$ sub-lattices
labelled by $\alpha$. We label the points on the $BZ$ by $l$ (an integer) and
denote the occupation number of the $(\mathbf k_l,\alpha)$ mode by
$n_{\mathbf k_l\alpha}$.  The collection of all the occupation numbers is
denoted by $\left\{n\right\}$.  The empty state ($n_{\mathbf
k_l\alpha}=0,~\forall\mathbf k_l,\alpha$) is denoted by $\vert 0\rangle$. The
Fock basis is,

\begin{equation}
\label{fbdef}
\vert\left\{ n\right\}\rangle=\prod_{l,\alpha}
\left(C^\dagger_{\mathbf k_l\alpha}\right)^{n_{\mathbf k_l}}\vert 0\rangle,~~
C^\dagger_{\mathbf k_l\alpha}C_{\mathbf k_l\alpha}\vert\left\{n\right\}\rangle
=n_{\mathbf k_l\alpha}\vert\left\{n\right\}\rangle
\end{equation}

where $(C^\dagger_{\mathbf k_l\alpha}, C_{\mathbf
k_l\alpha}),~\alpha=1,\dots,N_B$ are the fermion creation and annihilation
operators.

Any many-body state, $\vert\psi\rangle$ can be expanded as,
\begin{equation}
\label{psiexp}
\vert\psi\rangle=\sum_{\{n\}}\psi(\{n\})\vert\{n\}\rangle.
\end{equation}
We define the exchange operators, $E(\mathbf k_i,\mathbf k_j)$, by their action
on the Fock basis. These operators exchange the occupation numbers of the modes
at $\mathbf k_i$ and $\mathbf k_j$. We define,
\begin{equation}
\label{ek1k2alphadef}
E(\mathbf k_i,\mathbf k_j)\vert \dots,n_{\mathbf k_i},\dots,
n_{\mathbf k_j},\dots\rangle\equiv
\vert ..,n_{\mathbf k_j},..,n_{\mathbf k_i},..\rangle.
\end{equation}
The quantum distance between $\mathbf k_i$ and $\mathbf k_j$ is then
defined as,
\begin{equation}
\label{dk1k2def}
d(\mathbf k_i,\mathbf k_j)\equiv\sqrt{1
-|\langle\psi\vert E(\mathbf k_i,\mathbf k_j)\vert\psi\rangle|^{\alpha}}.
\end{equation}

The above definition satisfies all the properties of distances including
triangle inequalities \cite{Paper1}.  For a $d$-dimensional lattice with $L^d$
number of total sites, with every $d$ dimensional vector $\mathbf k_i = (k_i^{1},k_i^{2},\dots k_i^{d})$ 
representing a point in the spectral parameter space $\mathbf k_i \in \mathbf k$,  we associate a 
unique integer $i$, which runs from $1$ to $L^d$. So the above distances defined in Eq.~\eqref{dk1k2def}, 
gives us a $ L^d \times L^d$  distance matrix $D$ whose elements are given by

\begin{equation}
\label{dmatdef}
D(i,j)\equiv\sqrt{1
-|\langle\psi\vert E(\mathbf k_i,\mathbf k_j)\vert\psi\rangle|^{\alpha}}.
\end{equation}

\subsection{The graph of a state}
\begin{figure}[h!]
\includegraphics[width=4cm]{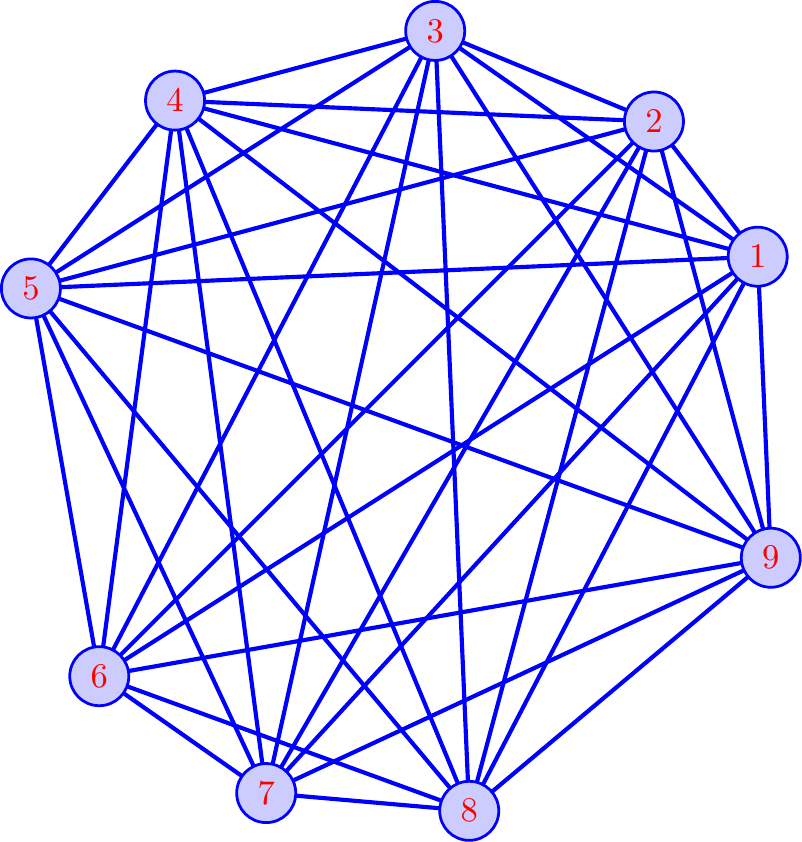}
\caption{The schematic representation of the graph of a 9-site system.}
\label{graph}
\end{figure}

In this paper, all our explicit computations are for a finite site system
detailed below. Thus we consider the problem of extracting geometric
information from a $L^d \times L^d$ distance matrix $D$, of a many-body state, 
defined between any two points on the set of quasi-momenta, denoted by $BZ$. We
refer to the combination, $G\equiv(BZ,D)$ as the graph of the state.  The
quasi-momenta are the vertices of the graph. Any pair of quasi-momenta define
an edge and $D$ defines the length, of each edge.  Fig.~(\ref{graph}) shows the
schematic figure for a graph of a 9-site system.

\subsection{The 1-dimensional $t$-$V$ model}
\label{1dtvm}

As mentioned above, we study a single band model on a one dimensional
lattice consisting of $L$ sites with nearest neighbour repulsive interactions,
the so called $t$-$V$ model. The Hamiltonian is
\begin{eqnarray}
H      &=&H_0+H_{int}\\
H_0    &=&-t\sum_{i=1}^{L}\left(C_i^\dagger C_{i+i}+H.c.\right)\\
H_{int}&=&V\sum_{i=1}^L~n_in_{i+1}
\end{eqnarray}
where $n_i\equiv C^\dagger_iC_i$ and we impose the periodic boundary
conditions, $C_{i+L}=C_i$. The (dimensionless) quasi-momenta can then be
chosen to be $q_k=\frac{2\pi}{L}k,~k=1,\dots,L$. The operators that create
and annihilate fermions with quasi-momentum $q_k$ are 
\begin{eqnarray}
C^\dagger_k&\equiv&\frac{1}{\sqrt L}\sum_{i=1}^L~e^{-iq_ki}C^\dagger_i\\
C_k&\equiv&\frac{1}{\sqrt L}\sum_{i=1}^L~e^{iq_ki}C_i.
\end{eqnarray}

In this one band model, for translationally invariant states,
the expression for the expectation values of the exchange operators in terms of
the fermion operators has been shown to be \cite{Paper1},
\begin{equation}
\label{en1n2}
\langle\psi\vert E(k_1,k_2)\vert\psi\rangle=1
-\langle\psi\vert\left(n_{k_1}-n_{k_2}\right)^2\vert\psi\rangle
\end{equation}
where $n_k\equiv C^\dagger_kC_k$.

As described in our previous paper \cite{Paper1}, we have solved for the 
ground state of this model numerically for $L\le 18$ and $V\le 12$.

\section{The intrinsic curvature}
\label{curvature}

In the thermodynamic limit, the $BZ$ is a continuous manifold (topologically, a 
torus). If the distance matrix $D(\mathbf k_1,\mathbf k_2)$ is a smooth 
function of its arguments, then local geometric objects like the metric
and curvature can be defined by standard methods.

For finite-site  correlated systems, the $BZ$ is a discrete and finite set.
We do not have a continuous manifold but only a graph as defined above.
Various definitions have been proposed for the notion of 
curvature for a graph \cite{Curvature,Forman,OLLIVIER2009,OC2}. 
In this work, we use the defintion of the curvature proposed by 
Ollivier, for a graph \cite{OC2,lin2011},  called the Ollivier-Ricci Curvature. 

\subsection{The Ollivier-Ricci curvature of a graph}
\label{orc}

The definition of the Ollivier-Ricci curvature for a graph is motivated
by the following definition of the Ricci curvature of a continuous
manifold \cite{ollivier:hal-00858008}.

Consider a smooth, $N$-dimensional Riemannian manifold $\mathcal{M}$. Denote the local
coordinates by $x^\mu,~\mu=1,\dots,N$ and the metric by $g_{\mu\nu}(\mathbf
x)$. Consider a nearby point with local coordinates  $\mathbf y=\mathbf
x+\delta\mathbf v$, where $\mathbf v$ is a unit tangent vector at $\mathbf x$.
Define two $\epsilon$-balls, $m_{x}^{\epsilon}$ and $m_{y}^{\epsilon}$ around $\mathbf x$ and
$\mathbf y$ to consist of the points in $\mathcal{M}$ that are at a distance $\le \epsilon$ from
$\mathbf x$ and $\mathbf y$ respectively. 

\begin{figure}[h!]
\centering
\includegraphics[angle=0,width=0.4\textwidth]{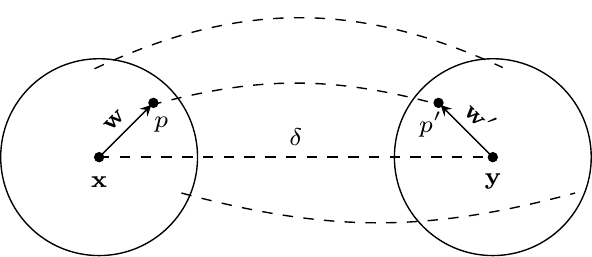}
\caption{Two nearby balls, $m_{x}^\epsilon$ and $m_{y}^\epsilon$ of radius $\epsilon$ whose centres
are a small distance, $\delta$, apart, along the unit vector $\mathbf v$.  
Parallel transport of a unit vector $\mathbf w$ gives $\mathbf w'$  such that for a
point $p\in m_{x}^\epsilon$ parallel transport along the geodesic of length $\delta$ 
yields a point $p^\prime\in m_y^\epsilon$. The average distance between the points $p$ and
$p^\prime$ in flat space is equal to $\delta$, while in the presence of 
curvature the lowest-order deviation from $\delta$ is given by 
Equation \eqref{contkappa}.} 
\label{manifold}
\end{figure}

It has been been shown  \cite{ollivier:hal-00858008} that if $m_{x}^{\epsilon}$ is
mapped to  $m_{y}^{\epsilon}$ using the Levi-Civita connection, then the average
distance between the points  $p\in m_{x}^{\epsilon}$  and their images
$p^\prime\in m_{y}^{\epsilon}$, in the limit $\delta, \epsilon\rightarrow 0$, is
\begin{equation}
\label{contkappa}
W(m_{x}^\epsilon, m_{y}^\epsilon)=\delta(1-\frac{1}{2(N+2)}
\kappa(\mathbf v,\mathbf v)+O(\epsilon^3 +\delta \epsilon^2)),
 \end{equation}
where $\kappa(\mathbf v, \mathbf v)$ is the Ricci curvature associated with 
the unit vector $\mathbf v$. This is schematically illustrated in Figure
(\ref{manifold}).

The above discussion motivates the definition of the curvature of graphs
as follows  \cite{ollivier:hal-00858008,lin2011}.
Replace the $\epsilon$-ball, $m_x^\epsilon$ around $\mathbf x$ by the normalised 
distribution of distances of all the vertices in the graph from the vertex $i$,
\begin{equation}
\label{mgraphdef}
 m_{i}(j)=\frac{D(i,j)}{\sum_k D(i,k)}. 
\end{equation}
Replace the average distance $W$ by the so called Wasserstein distance, 
$W(m_i,m_j)$, defined as the distance between two distributions, $m_i$ and
$m_j$ as follows \cite{Villani}:
\begin{equation} 
\label{Wdef} 
W(m_i,m_j)\equiv  \inf_{\Pi}\sum_{kl} D(k,l)\Pi_{ij}(k,l)
\end{equation}
where $\Pi_{ij}(k,l)$ is a joint probability distribution defined by,
\begin{equation}
\label{pidef}
\sum_l\Pi_{ij}(k,l)=m_i(k),~~\sum_k\Pi_{ij}(k,l) =m_j(l).
\end{equation}

The curvature, $\kappa(i,j)$, corresponding an edge  $(i,j)$ of the graph
is then defined by, 
 \begin{equation}
\label{kappaijdef}
W(m_i,m_j)=D(i,j)(1-\kappa(i,j)).
\end{equation}

The corresponding generalization of the scalar curvature at a point $\mathbf x$
in Riemannian manifold for a graph is \cite{Jost:2014:ORC}
\begin{equation}
\label{kappaidef}
 \kappa(i)= \frac{1}{\sum_{k}D(i,k)} \sum_j \kappa(i,j).
\end{equation}

\subsection{Numerical results for the intrinsic curvature}

We have numerically computed the ground state of the Hamiltonian of the 
1-dimensional $t$-$V$ model defined in Section \ref{1dtvm} for $L=18$. 
We have then computed the $18\times18$ distance matrix, $D(i,j)$, as defined 
by Equations (\ref{ek1k2alphadef}) and (\ref{dk1k2def}). We have choosen 
$\alpha=2$ in Eq.~\eqref{dmatdef} to obtain the distance matrix. 

The problem of computing the Wasserstein distance, $W(m_i,m_j)$, is a problem in
linear programming \cite{OR_LP}. We have to minimize the linear function of
$\pi_{ij}$ given in Equation~\eqref{Wdef}, subject to linear constraints given
in Equation~\eqref{pidef}. We do this numerically, using the standard techniques
of linear programming.

The curvature along each edge of the graph defined in Equation 
(\ref{kappaijdef}) and the scalar curvature defined in Equation 
(\ref{kappaidef}) are then computed.

To characterise the values of curvatures along the edges in the metallic state, 
we find it convenient to classify the edges into two types, as follows.
The non-interacting system defines a Fermi sea of occupied single particle
states. We denote the quasi-momenta of the occupied states by $k_{in}$
and the quasi-momenta of the unoccupied states by $k_{out}$.

\begin{figure}[h!]
\centering
\includegraphics[angle=0,width=0.45\textwidth]{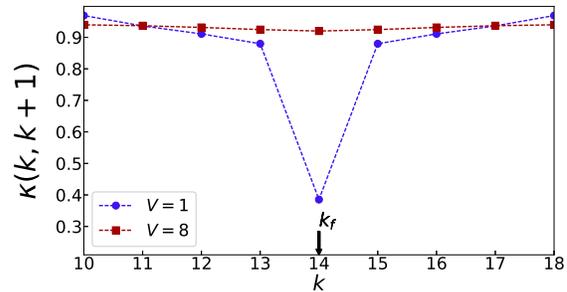}
\caption{Curvatures for the nearest neighbour edges ($k,k+1$) over half the 
$BZ$, where $k$ are the quasi-momenta modes corresponding to the vertices of 
the graph, for different interaction strengths. The metallic regime is 
characterised by a discontinuity at the Fermi point $k_f$.}
\label{OR-1}
\end{figure}

\begin{figure}[h!]
\centering
\includegraphics[angle=0,width=0.45\textwidth]{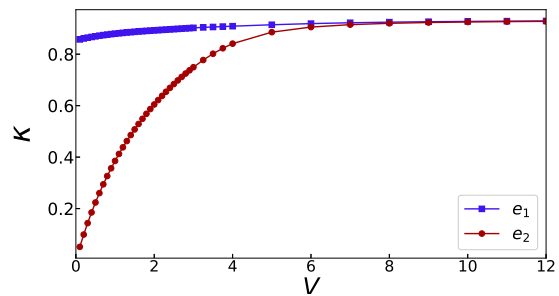}
\caption{Curvatures for both type of edges $e_1$ and $e_2$ as function of interaction
strength $V$.}
\label{OR-2}
\end{figure}

In the metallic regime, $V\ll 2$, we find that the curvatures classify the edges
into two classes, $e_{1} \equiv (k_{in},k_{in})$ or ($k_{out},k_{out}$) and
$e_{2}\equiv (k_{in},k_{out})$ or ($k_{out},k_{in}$). The edges $e_1$ have large
curvatures whereas the edges $e_2$ have small curvatures. On the other hand the
curvatures of $e_1$ and $e_2$ are both quite large and uniform in the
insulating regime, $V\gg 2$. This is illustrated in Figure (\ref{OR-1}).
Fig.~(\ref{OR-2}) shows the Ricci curvature of both type of edges as a
function of $V$. The curvature of $e_1$ is more or less constant at all the 
values of the  interaction. However, in the metallic and the cross over
regime, the curvature of $e_2$ is continously changing. After the cross
over it saturates to a  constant value, approximately the same as $e_1$.

\begin{figure}[h!]
\includegraphics[angle=0,width=0.39\textwidth]{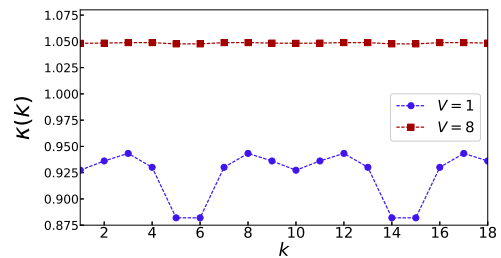}
\caption{Scalar Curvature as a function of the quasi-momenta modes representing vertices
of the graph. In insulating regime the scalar curvature is uniform over all the vertices.}
\label{OR-scalar}
\end{figure}

Fig.~(\ref{OR-1}) shows the plot of curvature for nearest neighbour edges
$\kappa (k,k+1)$ in the two regimes.  It can be seen for $V=1$, in the metallic regime 
there is a discontinuity  in the curvature, at the Fermi point which is labelled by $k_{f}$. 
This discontinuity decreases as function of interaction strength and  vanishes
at large $V$. The insulating regime is characterised by a uniform scalar curvature, 
whereas the scalar curvature varies considerably for different vertices in the metallic phase. 
This is illustrated in Fig.~(\ref{OR-scalar}) which shows the plot of the scalar curvature as 
a function of the quasi-momenta modes.

Thus the ``clustering'' of the distances in metallic phase that we find in 
our previous study \cite{Paper1}  manifests in the Ricci curvature by distinct differences  
in the curvature along the two different types of edges, $e_1$ and $e_2$.
We also find a variable scalar curvature in this phase. The ``declustered'' uniform distances in 
the insulating regime found in our previous work \cite{Paper1} manifests as 
homogenous curvatures.

\section{Isometric embedding of the distance matrix in a Euclidean space}
\label{ee}

In this section we explore the extrinsic geometry of the distance matrix by
isometrically embedding it in a Euclidean space. Namely, we attempt to find
a set of points, $\mathbf x_i$, in some Euclidean space such that,
\begin{equation}
\label{eeded}
D(i,j)=\vert\mathbf x_i-\mathbf x_j\vert.
\end{equation}

First, we show that the distance matrices of mean field
states can always be embedded in a finite dimensional Euclidean space. 
Then, we examine the general method of isometrically embedding 
a distance matrix in a Euclidean space. We follow this by exact analytical 
results for the extreme values of the coupling, namely $V=0,\infty$. 
Finally, we apply the method to the numerically computed distance matrix of 
the 18-site $t$-$V$ model.

\subsection{Embedding of mean field states}
\label{mfsee}
In previous work \cite{Paper1} we had shown that for a tight binding
model with one band completely filled and the others completely empty, our
definition of the quantum distance matrix, in terms of the expectation values
of the exchange operators reduces to the standard definition in terms of
one particle wavefunctions. In Appendix \ref{mfdm}, we generalise this result
to the case of a general mean field state with several filled and partially 
filled bands. In 
particular, we show that for the case of the mean field states of an 
$N_B$ band tight binding model, with $N_F(\mathbf k)$ of the bands occupied at 
$\mathbf k\in BZ$, the distance matrix is,
\begin{eqnarray}
\nonumber
D_{\alpha}(\mathbf k_1,\mathbf k_2)&=&\sqrt{1-
\vert\langle\Psi(\mathbf k_1)\vert\Psi(\mathbf k_2)\rangle\vert^{2\alpha}
\delta_{N_F(\mathbf k_1),N_F(\mathbf k_2)}}
\end{eqnarray}
where $\vert\Psi(\mathbf k)\rangle$ is the Slater determinant 
of  the occupied single particle wave functions at $\mathbf k$. 

In this section, we show that, at $\alpha=1$, namely the Hilbert-Schmidt
distance matrix, in a finite dimensional Hilbert space can always be 
isometrically embedded in a finite dimensional Euclidean space. We further
show that for arbitrary $\alpha$, while the embedding in a finite 
dimensional Euclidean space is not isometric, the metric, which gives the 
distance between two infinitesimally nearby points, is the same for all
$\alpha$ up to a scaling factor.

We consider a $N$-dimensional Hilbert space, ${\cal H}_N$.
The matrix of Hilbert-Schmidt distances between 
$\vert\Psi_n\rangle,~\vert\Psi_m\rangle\in {\cal H}_N$, is
\begin{equation}
\label{qdmdef}
(D(n,m))^{2}=1-{\rm tr}\left(\rho_n\rho_m\right)
\end{equation}
where $\rho_n\equiv\vert\Psi_n\rangle\langle\Psi_n\vert$.

Any linear Hermitian operator in ${\cal H}_N$ can
be expressed as a linear combination of the $N^2$ generators of $U(N)$ in
the fundamental representation. We can always take these to be the identity
operator, $I$, and the generators of $SU(N)$,
$T^\alpha,~\alpha=1,\dots,N^2-1$. These can be chosen to satisfy the
conditions
\begin{equation}
\left(T^\alpha\right)^\dagger=T^\alpha,~~~{\rm tr}\left(T^\alpha\right)=0,
~~~\frac{1}{2}{\rm tr}\left(T^\alpha T^\beta\right)=\delta^{\alpha\beta}.
\end{equation}

We can express $\rho_n$ as,
\begin{equation}
\rho_n=\frac{1}{N}+\vec a_n\cdot\vec T,
~~~~~a_n^\alpha=\frac{1}{2}{\rm tr}\left(T^\alpha\rho_n\right).
\end{equation}
Since $T^\alpha$ and $\rho_n$ are Hermitian matrices, $\vec a$ is a real
$N^2-1$ dimensional vector, i.e  $\vec a_n\in \Bbb R^{(N^2-1)}$.

 The fact that $\rho^2_n=\rho_n$ and ${\rm tr}\rho_n=1$ implies that
\begin{equation}
\vec a_n\cdot\vec a_n=\frac{1}{2}\left(1-\frac{1}{N}\right).
\end{equation}
Note that $\rho_n^2=\rho_n$ implies other constraints on $\vec a_n$ as well
but these are not relevant for the current proof.

The square of the above matrix is computed to be,
\begin{eqnarray}
(D(n,m))^{2}&=&1-\frac{1}{N}-2\vec a_n\cdot\vec a_m\\
                             &=&\vert\vec a_n-\vec a_m\vert^2.
\end{eqnarray}
Thus we have shown that the Hilbert-Schmidt distance matrix of mean field 
states can be isometrically embedded in a finite dimensional Euclidean space. 
Note that this result is true for a system with arbitrary number of sites and 
hence holds in the thermodynamic limit.

Let us now consider the limit of $N\rightarrow\infty$, $2\pi{\mathbf n}/N
\rightarrow \mathbf k$, where we have taken the index $n$ to represent a 
set of integers, $\mathbf n$.
For arbitrary $\alpha$ the distances between neighbouring points
$\mathbf k$ and $\mathbf k+d\mathbf k$, in the limit $d\mathbf k\rightarrow 0$ is,
\begin{eqnarray*}
\left(D_\alpha(\mathbf k,\mathbf k+d\mathbf k)\right)^2&=&1-
\left({\rm tr}\rho(\mathbf k)\rho(\mathbf k+d\mathbf k)\right)^\alpha\\
&=&1-\left(1-\vert\vec a(\mathbf k+d\mathbf k)-\vec a(\mathbf k)
\vert^2\right)^\alpha.
\end{eqnarray*}
If the wave functions, $\vert\Psi(\mathbf k)\rangle$, are smooth functions of 
$\mathbf k$, so will be $\vec a(\mathbf k)$. We then have,
\begin{eqnarray}
\left(D_\alpha(\mathbf k,\mathbf k+d\mathbf k)\right)^2&=&
\alpha\frac{\partial \vec a(\mathbf k)}{\partial k_\mu}\cdot
\frac{\partial\vec a(\mathbf k)}{\partial k_\nu}dk_\mu dk_\nu\\
&\equiv&g_\alpha^{\mu\nu}(\mathbf k)dk_\mu dk_\nu.
\end{eqnarray}
This implies that
\begin{equation}
g_\alpha^{\mu\nu}(\mathbf k)=\alpha g_1^{\mu\nu}(\mathbf k).
\end{equation}

To conclude, we have shown: (a)
for $\alpha=1$, mean field states corresponding to a finite number of bands
can be isometrically embedded in a finite dimensional Euclidean space,
(b) for $\alpha\ne 1$, this embedding does not preserve the distances, however,
in the thermodynamic limit, the distances between neigbouring points in the 
spectral parameters are just scaled by the factor $\alpha$. Thus the shape
of the embedded surface is independent of $\alpha$ up to a scaling factor.

\subsection{Isometric Euclidean embedding of a general distance matrix}
\label{genee}

The problem of isometrically embedding a distance matrix in a Euclidean
space is a well studied one \cite{Schoenberg,DG,DG2}. In this section we review 
the general method of doing so.

Consider a general $L^{d}\times L^{d}$ distance matrix, $D(i,j),~i,j=1,\dots, L^{d}$.
Construct the so called Gram matrix,
\begin{equation}
\label{gmatdef}
G =-\frac{1}{2} (I-\frac{1}{L^d}ee^{T}) D^{2} (I-\frac{1}{L^d}ee^{T})
\end{equation}
where, $D^2$ is the matrix of squared distances, $D^2(i,j)\equiv (D(i,j))^2$.
$I$ is the $L^d\times L^d$ identity matrix. $e$ is a $L^d$-dimensional column vector
with all entries equal to 1.

It has been proved \cite{Schoenberg} that $D$ is the distance matrix in a $N_d$
dimensional Euclidean space if and only if the matrix $G$ is a positive
semi-definite matrix of rank $N_d$. Further, if this is so, then $G$ can always
be written as,
\begin{equation}
G(i,j)=\sum_{n=1}^{N_d}x^n_ix^n_j
\end{equation}
where ${\mathbf x}_i$ are $N_d$ component column vectors, with components 
$x^n_i,~n=1,\dots,N_d$ and 
\begin{equation}
D^2(i,j)=\sum_{n=1}^{N_d} (x_i^n-x_j^n)^2.
\end{equation}

\subsection{Isometric embedding at the extreme limits}
\label{edmextremes}

In this section, we study the problem of isometrically embedding the distance
matrix of the ground state of the half filled, 1-dimensional $t$-$V$ model at
$V=0$ and $V=\infty$.  These limits are simple and we have analytically
computed the distance matrices for this model in our previous
paper \cite{Paper1}. These results can easily be generalised to a
$d$-dimensional model. However, we postpone that for a future work, since in
this paper our numerical results at finite, non-zero $V$ are only for the
1-dimensional model.

\subsubsection{$V=0$}

$V=0$ is the non-interacting model. The ground state is a mean field state (the
Fermi sea) with all the single particle states with energies less than zero
occupied and those with energies greater than zero unoccupied. We denote the
spectral parameters of the occupied single particle states by $k_{in}$
and the spectral parameters of the unoccupied single particle states by
$k_{out}$. We have shown in previous work \cite{Paper1}, the distance
matrix, denoted by $D_{FS}$ is,
\begin{eqnarray}
D_{FS}(k_{in},k_{in})&=0=&D_{FS}(k_{out},k_{out})\\
D_{FS}(k_{in},k_{out})&=1=&D_{FS}(k_{out},k_{in}).
\end{eqnarray}
It trivially follows that this distance matrix can be embedded in a one
dimensional Euclidean space,
\begin{equation}
x(k_{in})= -0.5,~~~~~x(k_{out})= 0.5.
\end{equation}

\subsubsection{$V=\infty$}

In this limit the ground state is doubly degenerate. It is a charge density
state (CDW) with either the odd sites occupied and the even sites empty or the
odd sites empty and the even sites occupied. We consider the translationally
invariant case which is the symmetric sum of these two states. In previous
work \cite{Paper1} we had analytically derived the distance matrix in this
limit to be,
\begin{eqnarray}
\label{dcdw1}
D^{CDW}_{ij} = \begin{cases} 
                 0 & i=j \\        
                 \frac{\sqrt 3}{2} & i\ne j, \, i \ne j+\frac{L}{2}\\              
                 1 &  i=j+\frac{L}{2}             
               \end{cases}
\end{eqnarray}
where $i,j=1,\dots,L$ label the points in the $BZ$, $k_i=2\pi i/L$.

It is useful to define the following two $(L/2)\times (L/2)$ matrices. 
$I$ is defined as the identity matrix and ${\cal I}$ is defined as the 
matrix with all entries equal to one. With these definitions, we can
write
\begin{equation}
\label{dcdwblocks}
D^{CDW}=\frac{\sqrt 3}{2}\left(\begin{array}{cc} 
{\cal I}-I&{\cal I}-I\\{\cal I}-I&{\cal I}-I\end{array}\right)
+\left(\begin{array}{cc}0&I\\I&0\end{array}\right).
\end{equation}

In Appendix \ref{cdwee} we have applied the procedure described in section
\ref{genee} to the above distance matrix. The simple structure of the distance
matrix makes the problem exactly solvable. We have shown that rank of the Gram
matrix is equal to $L-1$. Thus, the dimension of the embedding Euclidean space 
is $L-1$, which grows as the volume of the system in the thermodynamic limit.
We have also presented the explicit solution for the embedded vectors.

\subsection{Embedding at finite coupling}
\label{edmfinite}

We have numerically implemented the procedure described in section \ref{genee}
for the distance matrices computed for the ground state of the 18-site
1-dimensional $t$-$V$ model.  We find that for any non-zero $V$, it is always
possible to isometrically embed the distance matrix in a Euclidean space of
dimension $17$. Namely, unlike mean field states (Section \ref{mfsee}), as 
soon as the interaction is turned on, the rank of the Gram matrix becomes 
thermodynamically large, i.e $L-1$ and remains so till $V=12$. In the previous 
section we have shown that this is also true at $V=\infty$.

Based on the above results, we conclude that it is probably always possible to 
isometrically embed the distance matrix in a Euclidean space. However, for 
correlated states, the dimension of the embedding Euclidean space diverges
as the system size. 

When the correlations are non-zero but small, i.e at small values of $V$, we
may expect the state to be not very different from the mean field state. In the
1-dimensional system that we are analysing we know that as soon as interactions
are turned on, the system goes from a Fermi liquid to a Luttinger liquid. Thus
the ground state is qualitatively different as soon as the interaction is
turned on. Nevertheless, it remains metallic till $V/t=2$. This motivates us to
investigate if the distance matrix can, in some precise sense, still be
approximately embedded in a finite dimensional Euclidean space in the metallic
regime. There are several methods for approximate embedding of a distance matrix in a
Euclidean space
\cite{SVD_book,Indyk04low-distortionembeddings,Rabinovich}.  We
investigate two such methods in the next section.

\section{Approximate Euclidean embedding of the distance matrix}
\label{approxD}
In this section we continue the discussion in the end of previous section 
and ask: can we characterise the metallic state by the fact that its distance
matrix can be approximately embedded in a finite dimensional Euclidean space
with small (suitably defined) error?

We consider two well studied methods of approximate embedding:
(a) approximate embedding by truncation of the Gram matrix spectrum 
\cite{SVD_book} and
(b) approximate embedding by method of average distortion 
\cite{Rabinovich,ABRAHAM2011}.

We show that in the metallic regime ($V\ll2$), the distance matrix can be
embedded in finite dimensional Euclidean spaces with small ($\ll12\%$) 
error or average distortion in above methods, whereas in the
insulating regime the error or average distortion is much larger.

\subsection{Dimensionality reduction by truncation of Gram matrix spectrum with error estimate}
\label{TE}

The rank of the Gram matrix corresponds to the dimension of the Euclidean space
in which the distance matrix can be embedded (Sec.~\ref{genee}).  From the 
eigenspectrum of a given matrix to extract a subspace which is much smaller but
retains most of the information is a well studied problem
\cite{SVD_book}.  One of the methods to project to lower
dimensional spaces is to put all but the highest $q$ eigenvalues of the Gram
matrix equal to zero. The approximate Gram matrix thus obtained has rank $q$
and it yields an approximate embedding of the distance matrix in a
$q$-dimensional Euclidean space. The procedure is detailed below.

The Gram matrix is a real symmetric $L\times L$ matrix. It can hence be 
diagonalised by an orthogonal transformation,
\begin{equation}
\label{gdiag}
G=U\Lambda U^T
\end{equation}
where $U$ is an orthogonal matrix and $\Lambda$ a diagonal matrix. We choose
a basis where, $\Lambda_{11}\geq \Lambda_{22}\geq \cdots \geq \Lambda_{LL}$.
We then define an approximate Gram matrix in the diagonal basis, 
$\tilde\Lambda$ by putting all but the highest $q$ diagonal entries to zero,
$\tilde\Lambda_{ii}=\Lambda_{ii},~i\le q$, $\tilde\Lambda_{ii}=0,
i>q$. The approximate Gram matrix with rank $q$ is then defined as,
\begin{equation}
\label{gtildef}
\tilde G\equiv U\tilde\Lambda U^T
\end{equation}
The truncation error associated with keeping $q$ largest eigenvalues is
defined as,
\begin{equation}
E(q)= 1- \sqrt{\frac{\sum_{i=1}^{q}\Lambda_{ii}}{\sum_{k=1}^{L}\Lambda_{kk}}}.
\end{equation}
\begin{figure}
\centering
\includegraphics[angle=0,width=0.45\textwidth]{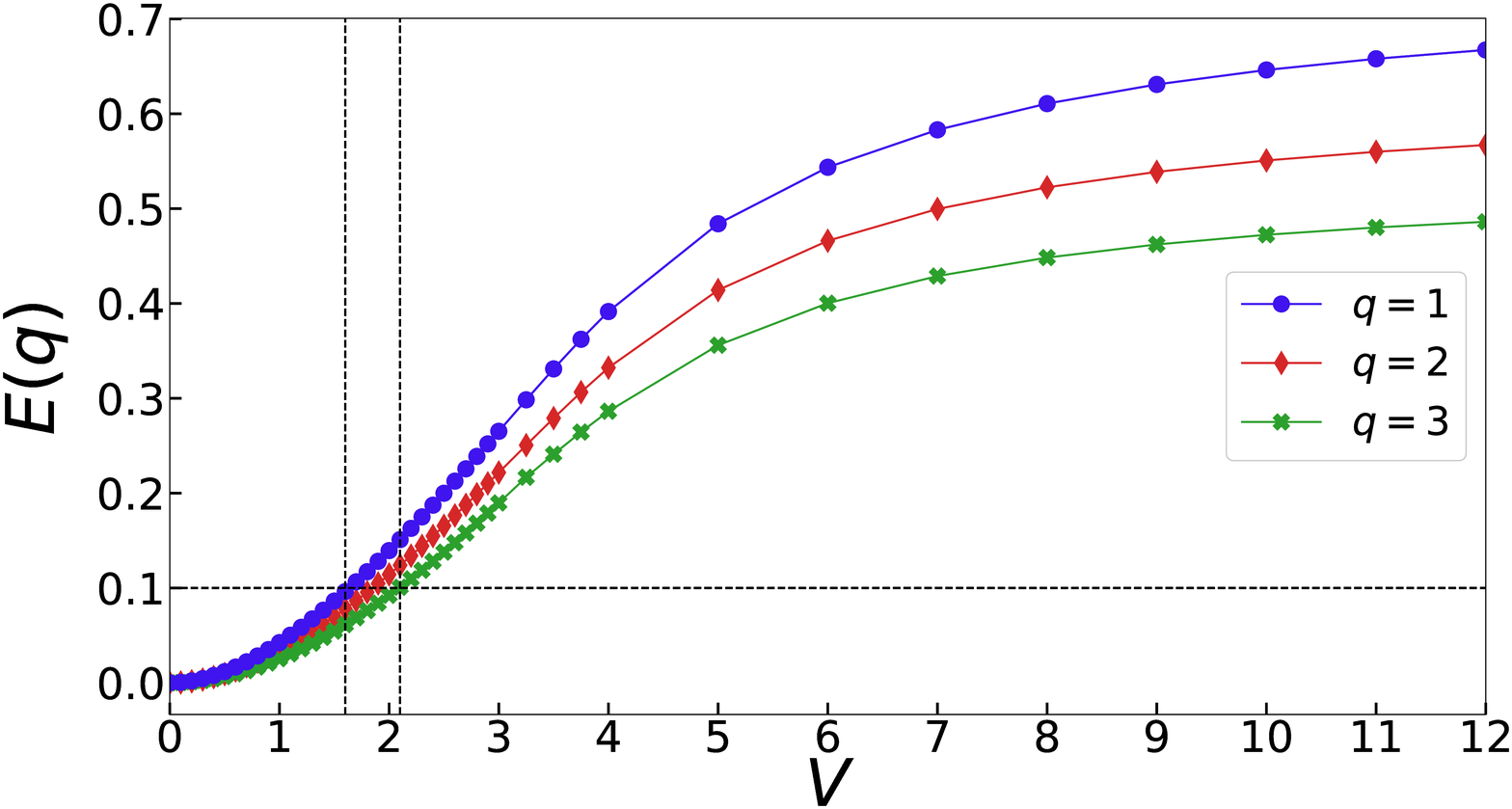}
\caption{Truncation error for keeping first few (1-3) eigenvalues for embedding $D$.
The truncation error for approximate embedding is less than $12\%$ in case of embedding
in one dimension $E(q=1)$ up to $V\approx 1.5$ and for embedding in three dimension $E(q=3)$ 
up to $V\approx 2$.}
\label{Error}
\end{figure}

Fig.~(\ref{Error}) shows how the truncation error for retaining first few
largest eigen values ($q=1-3$) behaves as a function of the interaction strength.
As can be seen in the figure, the truncation error is less than $12\%$, for
$q=1$ up to $V\approx 1.5$ and for $q=3$ up to $V\approx 2$.

\subsection{Embedding with distortion}
\label{D}

Another way of approximate embedding is the concept of embedding with
distortion \cite{Indyk04low-distortionembeddings,Rabinovich,ABRAHAM2011}. 
In general the method involves embedding the points in a low dimensional 
Euclidean space but with an error in the distance matrix. A measure of the 
distortion of the distance matrix, the average distortion, is defined as \cite{Rabinovich,ABRAHAM2011},
\begin{equation}
\label{avdistdef}
\epsilon_{avg}\equiv\frac{\sum_{i,j}D_{ij}}{\sum_{i,j}\tilde D_{ij}}
\end{equation}
where $\tilde D_{ij}$ are the distances in the low dimensional Euclidean
space. The low dimensional space is chosen such that the 
average distortion is minimized.

We implement this method as follows. We have a set of $L$ vectors in $L-1$ dimensions,
$\{\mathbf{x_i}\}$ ($i=1,\dots,L$), $\mathbf{x_i} \in \Bbb R ^{(L-1)}$,
obtained from isometric Euclidean embedding as discussed in Section \ref{edmfinite}. 
We explore the $(L-1)C_q$, $q$ dimensional subspaces obtained by picking $q$ of the basis
vectors, compute the average distortion for each case and pick the one that
minimizes it. We label the above minimum value of average distortion as $\epsilon_{min}$.
Note that our procedure is not optimal since we would probably
get lower values of the average distortion by rotating each set of basis. 
So what we obtain are upper bounds on the average distortion.

\begin{figure}[h!]
\centering
\includegraphics[angle=0,width=0.45\textwidth]{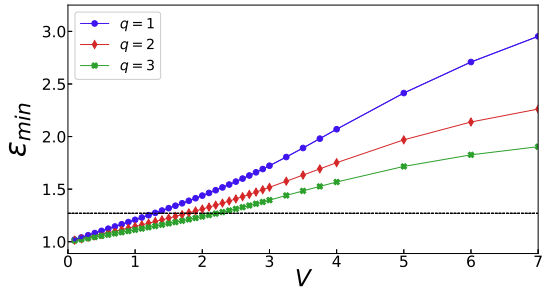}
\caption{Average distortion for approximate embedding of $D$ in lower 
dimensions as a function of interaction strength.}
\label{Distortion}
\end{figure}

Fig.~(\ref{Distortion}) shows the average distortion as a function of
interaction strength for $q=1,2,3$. Note that $\epsilon_{avg}=1$ corresponds to
the case where there is no distortion.  We find embedding with values of
average distortion very close to the value one is possible for small values of
interaction, $V \ll 2$.  Corresponding to the upper bound of error $0.12$ in
region $V\precsim2$ we find the maximum average distortion allowed to be
1.27. For $V>2$, the average distortion is large and the truncation error for
keeping three eigenvalues is high as well, so approximate Euclidean embedding
of $D$ into lower dimensional subspaces is not possible. 

\section{Approximate embedding of the Wasserstein distance}
\label{edmw}

In Section \ref{orc}, we had defined another distance function, $W(m_i,m_j)$,
namely the Wasserstein distance or the transportation distance associated with two
vertices of the graph. $W(m_i,m_j)$ is computed in terms of $D(i,j)$ and hence,
in principle, contains no information about the state. However, as we detail
in this section, the approximate embedding properties of the Wasserstein 
distance matrix seem to be more physically revealing than those of $D(i,j)$.

\begin{figure}[h!]
\centering
\includegraphics[angle=0,width=0.45\textwidth]{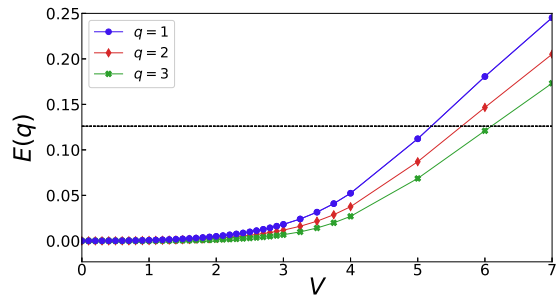}
\caption{Truncation error for keeping first few $(1-3)$ eigenvalues of $G$
as a function of the interaction strength, for approximate embedding of $W$.}
\label{Error-W}
\end{figure}

\begin{figure}[h!]
\includegraphics[angle=0,width=0.45\textwidth]{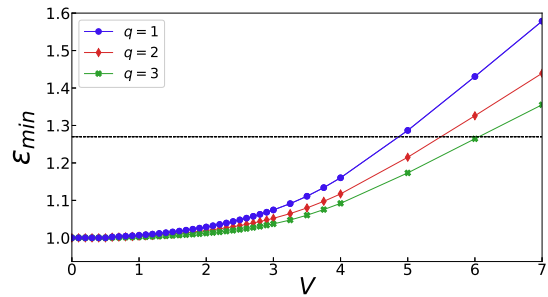}
\caption{Average distortion for approximate embedding of $W$
as a function of the interaction strength.}
\label{Distortion-W}
\end{figure}

In particular, we show that: (a) the embedding of $W(m_i,m_j)$ distinguishes 
between the metallic and insulating regimes more sharply than the embedding of 
$D(i,j)$. (b) $W(m_i,m_j)$ can be embedded in a finite dimensional Euclidean space 
with smaller error and average distortion than $D(i,j)$, for values of $V/t\gg2$, 
namely well in the insulating regime. 

Thus, $W(m_i,m_j)$ can be used to visualise the embedding in
the metallic as well as insulating regimes. We illustrate this by 
presenting the ``shapes'' given by the vector configurations 
obtained by low average distortion in both these regimes.

\subsection{Approximate embedding of $W$ by truncation}

\begin{figure*}[ht!]
  \begin{center}
   \includegraphics[angle=0,width=0.45\textwidth]{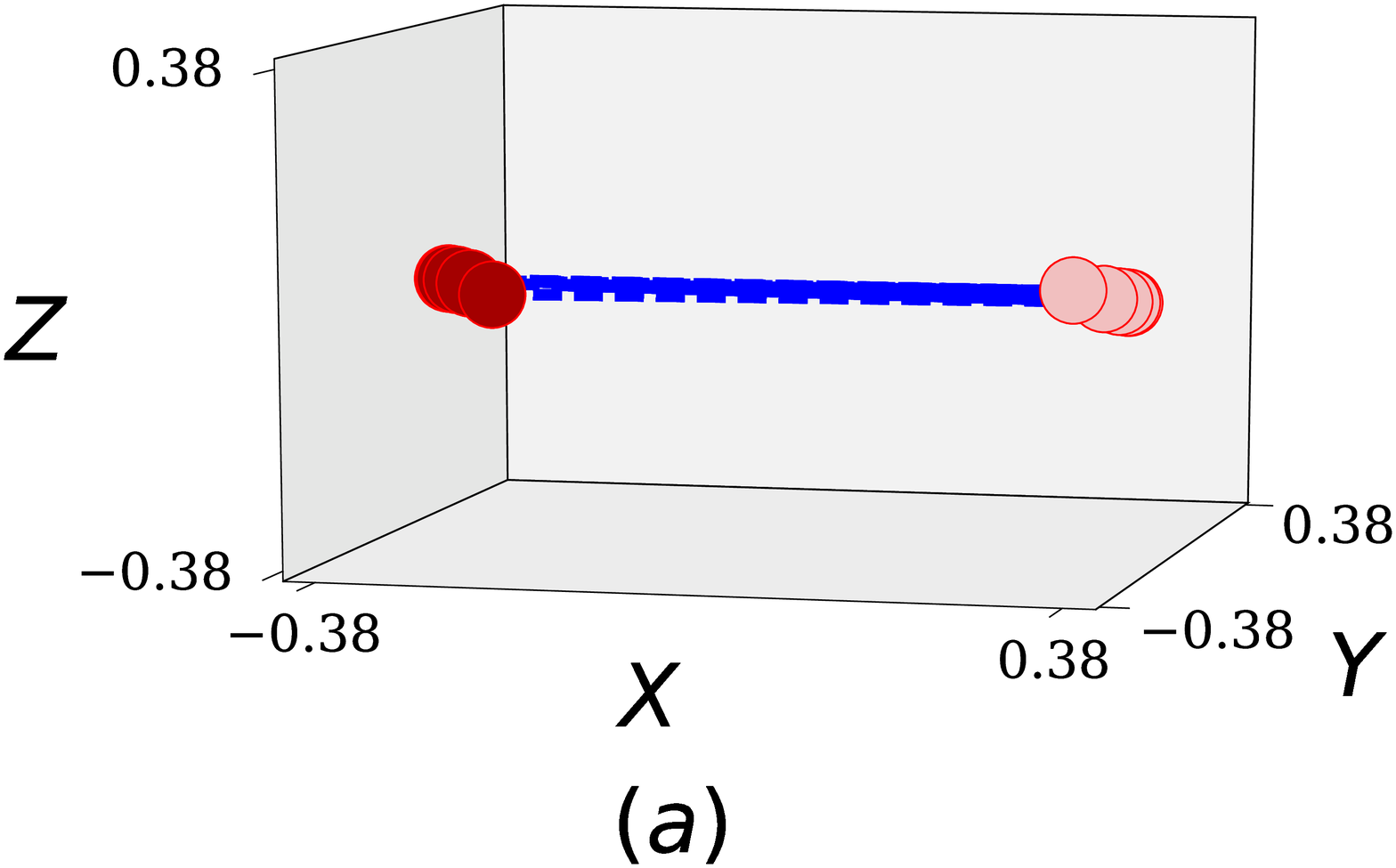}
   \includegraphics[angle=0,width=0.45\textwidth]{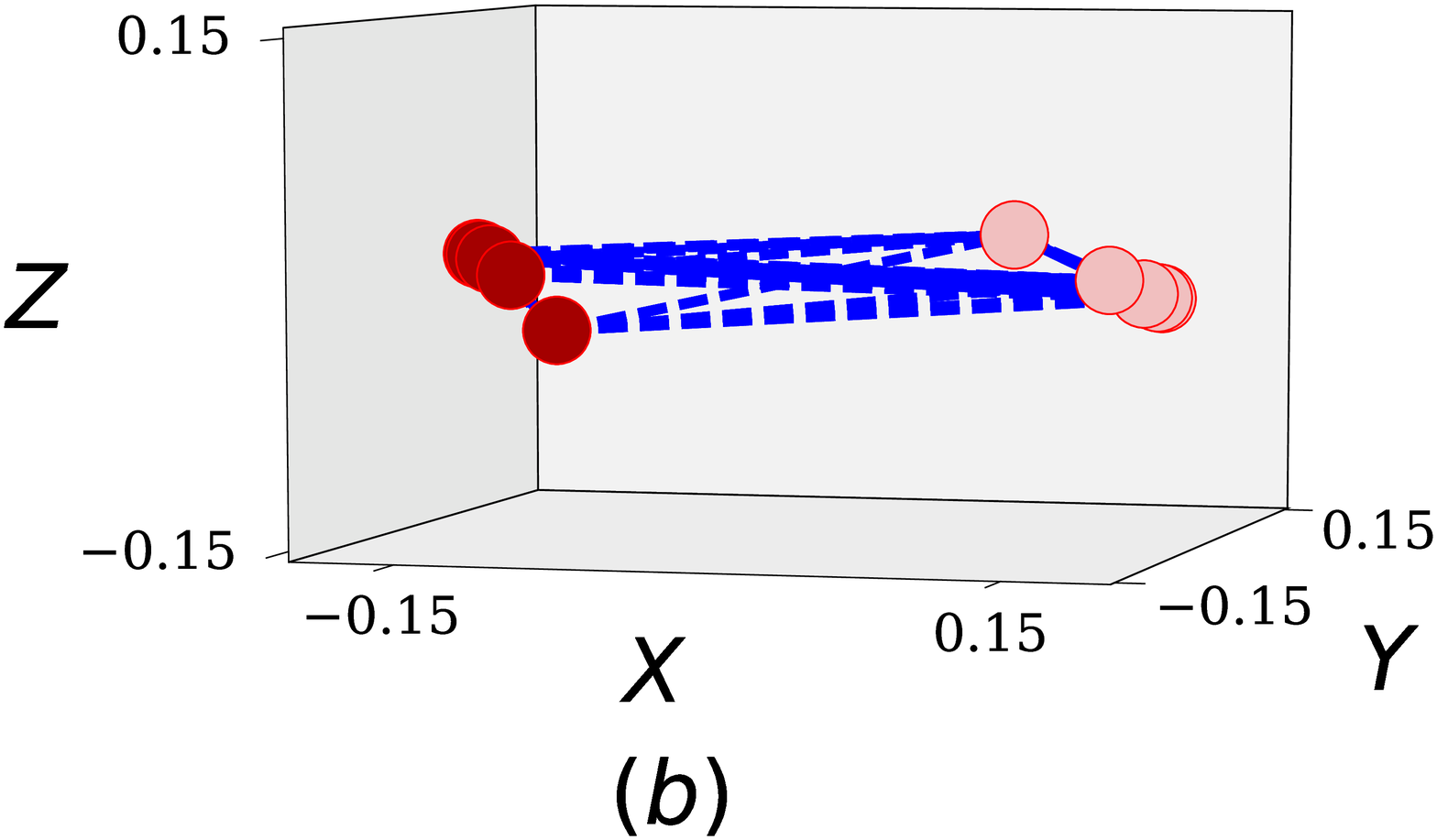}\\
   \includegraphics[angle=0,width=0.45\textwidth]{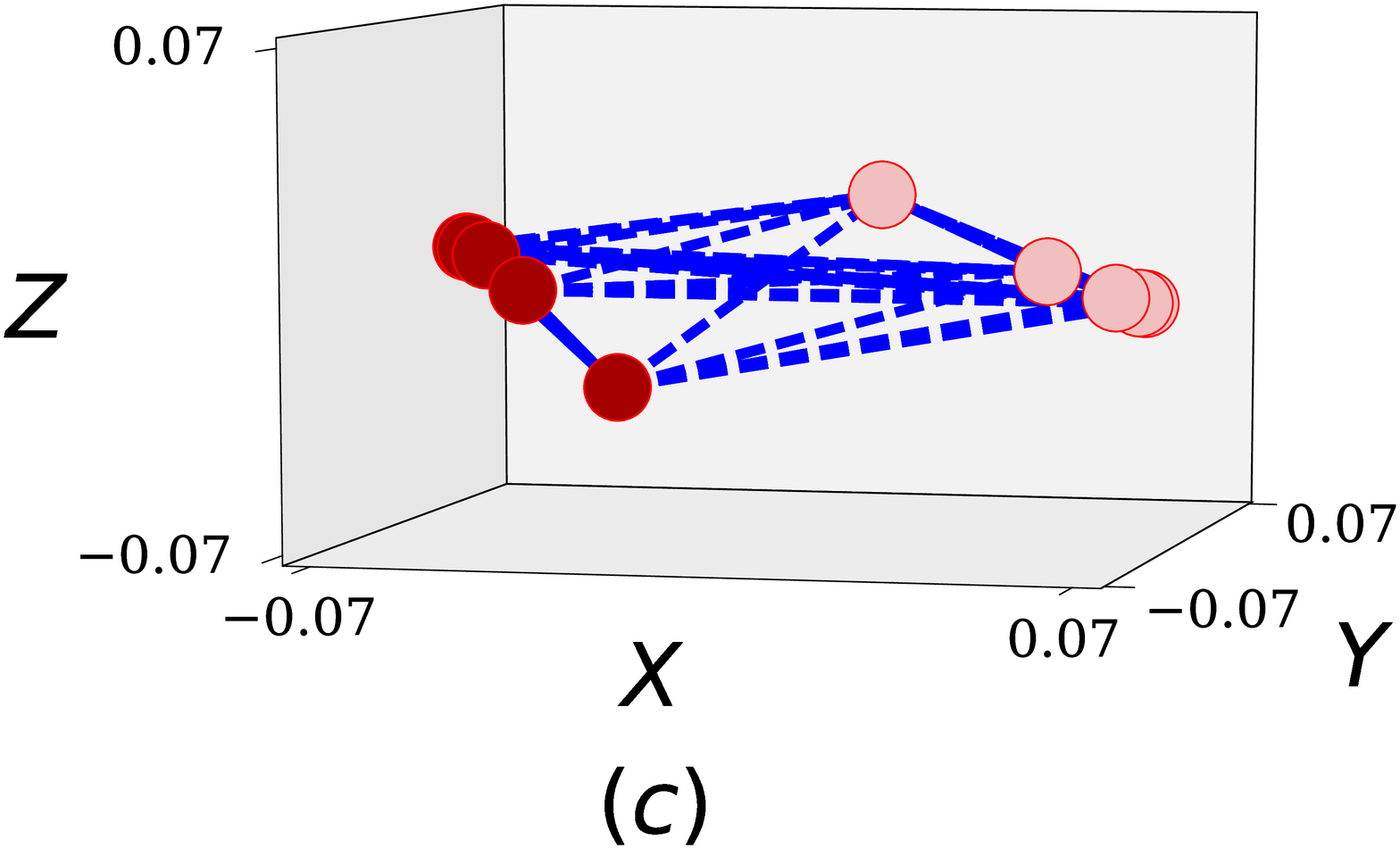}
   \caption{The embedded vectors of the Wasserstein distance in 
three dimensions. (a) $V=1$ with average distortion $<1\%$. (b) $V=3$ with
$3\%$ average distortion and (c) $V=6$ with $27\%$ average distortion.
Note that the scales of the axes are different in the three plots.}
  \label{Shape-W}
  \end{center}
\end{figure*}

At $V=0$, both the distance matrices $D$ and $W$ are identical, thus the
embedding gives two mirror points $k_{in}$ and $k_{out}$ in one-dimension
(Section \ref{edmextremes}).  Following the procedure of Section
(\ref{TE}) we compute the truncation error for keeping first three eigenvalues 
of $G$. The results are plotted in Fig.~(\ref{Error-W}). In the 
metallic regime, the error is extremely small even if only one eigenvalue 
is retained. The errors grow rapidly in the crossover regime and continue to
grow in the insulating regime.

\subsection{Approximate embedding of $W$ by distortion}

The average distortion for the embedding in one, two and three dimensions 
is plotted in  Fig.~(\ref{Distortion-W}). Again we see that the average 
distortion is almost negligible in the metallic regime and grows rapidly
after $V\approx 3$. However, well into the insulating regime, up to 
around $V=6$, the distortion remains reasonably small, less than about $27\%$. 

The above discussion shows that the Wasserstein distance matrix can be embedded
in one dimension with negligible distortion in the metallic regime and in low
dimensional Euclidean spaces well into the insulating regime with low
distortion. Thus, it seems to capture the physics of the state in a clearer
way than the distance matrix. It also seems to provide a way to visualise the 
many-body correlated state, both in the metallic and in the insulating 
regime. 

In Fig.~(\ref{Shape-W}), we have plotted the embedded vectors for
$V=1,3~{\rm and}~6$, for the three dimensional embedding. 
The filled circles represent $k_{in}$ and the unshaded 
circles represent $k_{out}$. Some points almost coincide and hence all
the $18$ points cannot be seen distincly in the figure.

Fig.~(\ref{Shape-W}-a) plots the embedded points at $V=1$ (metallic regime) with
very small ($<1\%$) average distortion. The two sets of points are clustered
around $(x,y,z)=(\pm 0.35,0,0)$. The range of the spread in the $y$ and $z$ coordinates are
$0.02$ and $0.03$ respectively. They are spread over a region less that $5\%$
of the range of the $x$ coordinate ($0.7$). Thus, set of embedded points, to a
very good approximation, lie in a one dimensional subspace.

Fig.~(\ref{Shape-W}-b) plots the embedded points at $V=3$ (crossover regime)
with $3\%$ average distortion. The points are all much closer to each other
but the relative spread in the $y$ and $z$ directions have increased. The 
ranges of the spread in the $x,y~{\rm and ~ z}$ coordinates are now $0.36$, $0.05$~and
~$0.06$ respectively. Thus the spreads in the $y$ and $z$ directions are now 
about $10\%$ of the spread in the $x$ direction.

The embedded points at $V=6$ with $27\%$ average distortion are plotted in
Fig.~(\ref{Shape-W}-c). The ranges of the spread in the $x,y$ and $z$ coordinates are
$0.13,0.04~{\rm and ~} 0.05$ respectively. The spreads in the $y$ and $z$
directions have now increased to $30\%$-$40\%$ of the spread in the $x$
direction.

The approximate embedding of the Wasserstein distances thus yields the following visualisation 
of the many-body state as a function of the interaction strength.  In the metallic regime 
the embedding basically consist of two points. One representing all the points in the Fermi 
sea and the other all the points outside it. In the crossover regime, the distances between 
the points reduce and hence all the points come closer to each other. The one-dimensional
nature is lost and the points spread out in the other two directions. This
trend continues in the insulating regime.

\section{Discussion and conclusion}
\label{concl}

To summarize our results, we have studied aspects of the intrinsic and
extrinsic geometry of the ground state of a correlated system, by analysing its
distance matrix on the Brillioun zone, defined in the previous work
\cite{Paper1}.

We have studied a system of interacting fermions on a finite size system. Hence
we have used the mathematical methods of discrete geometry to analyse our 
results.

First we have studied the intrinsic curvature, as defined by Ollivier
\cite{OLLIVIER2009,OC2} and have shown that this quantity is distinctly
different in the metallic and insulating regimes. The metallic regime is
characterised by non-uniform curvatures while insulating regime is homogenous,
characterised by  uniform curvatures.

We have then studied the extrinsic geometry of the state by analysing the exact
and approximate embedding of the distance matrix in Euclidean spaces.
The exact embedding at extreme limits of interaction reveals for the metal at $V=0$,
the distance matrix can be isometrically embedded in one dimension whereas for the
CDW insulator at $V=\infty$, the isometric embedding of the distance matrix
corresponds to an embedding dimension which scales as the system size and hence
is not finite in the thermodynamic limit.

We showed that the distance matrix can be embedded in a finite dimensional
Euclidean space with small error or average distortion in the metallic regime. 
This is not possible however in the insulating regime. 

We find that the Wasserstein distance matrix  constructed from the distance
matrix, can be embedded in a one dimensional space in the metallic regime.
Further, well within the insulating regime, it can be embedded in a finite
dimensional Euclidean space with relatively small error and average distortion.

It is very appealing to be able to characterise a correlated many-body by a 
surface in a finite dimensional Euclidean space since we have a good feeling
for Euclidean spaces. Our results indicate that while this is always possible
for mean field states, the dimension of the embedding space for the correlated states
may diverge as the system size, $L^d$. Since there are only $L^d$ embedded points, these
will not form any smooth surface. Methods of approximate embedding seem to
provide a method to obtain a smooth (though approximate) surface in a finite
dimensional Euclidean space for correlated states. In particular, the 
Wasserstein distance matrix, defined in terms of the distance matrix seems
to be more suited for this purpose rather than the distance matrix itself. 
We will be reporting on a more detailed analysis of this issue in a 
forthcoming paper.

\section{Acknowledgements}
We are grateful  to  R.~Simon, S.~Ghosh, R.~Anishetty, G.~Date and  
Emil Saucan for  useful  discussions.

\appendix

\section{Quantum distances of mean field states}
\label{mfdm}

In this section we generalise our previous results \cite {Paper1} of
the distance matrix for mean field states (MFS).

We consider a general $d$-dimensional lattice with $L$
unit cells in each direction. We label the sites of the unit cells by $i$ and the
sublattices by $a=1, \dots N_B$. The sites of the lattice are denoted by
$\mathbf R_{ia}=\mathbf R_i+\mathbf r_a$. Thus $\mathbf R_i$ specifies a point
in the unit cell and $\mathbf r_a$ the locations of the sub-lattice sites with
respect to that point. The fermion creation and annihilation operators are
denoted by $(C^\dagger_{ia}, C_{ia})$.  They satisfy the canonical
anti-commutation relations. The Fourier transforms of these operators are
defined as,
\begin{equation}
C_a(\mathbf k)\equiv\frac{1}{L^{\frac{d}{2}}}\sum_i
~e^{i\mathbf k\cdot\mathbf R_{ia}}C_{ia}
\end{equation}
where $\mathbf k\in BZ$.

We denote the single-particle hamiltonian in the quasi-momentum space by
$h_{ab}(\mathbf k)$ and its spectrum by,
\begin{equation}
h_{ab}(\mathbf k)u^n_b(\mathbf k)=\epsilon^n(\mathbf k)u^n_a(\mathbf k).
\end{equation}
We denote the Fermi level by $\epsilon_F$ and the number of occupied bands 
at $\mathbf k$ by $N_F(\mathbf k)$, namely,
\begin{equation}
N_F(\mathbf k)\equiv\sum_{n=1}^{N_B}
\Theta\left(\epsilon_F-\epsilon^n(\mathbf k)\right).
\end{equation}
The general mean field state is defined as
\begin{equation}
\label{genmfs1}
\vert u,\epsilon_F\rangle\equiv
\prod_{\mathbf k}\prod_{n=1}^{N_F(\mathbf k)}
\left(u^n_a(\mathbf k)C^\dagger_a(\mathbf k)\right)\vert 0\rangle
\end{equation}
where $u$ denotes the full set of eigenstates, $u^n(\mathbf k)$.
We define $\Psi_{a_1,\dots a_{N(\mathbf k)}}(\mathbf k)$ to be the 
antisymmetrised product of the $N_F(\mathbf k)$ single particle
wave functions, $u^n(\mathbf k)$,
\begin{equation}
\Psi_{a_1\dots a_{N_F(\mathbf k)}}(\mathbf k)=
\sum_P(-1)^P \prod_{l=1}^{N_F(\mathbf k)}u^n(\mathbf k)_{a_P(l)}
\end{equation}
The general mean field state can be written in the factorised
form,
\begin{eqnarray}
\nonumber
\vert\Psi(\mathbf k)\rangle&\equiv&
\left(\sum_a\Psi_{a_1\dots a_{N_F(\mathbf k)}}(\mathbf k)
\prod_{l=1}^{N_F(\mathbf k)}C^\dagger_{a_l}(\mathbf k)
\right)\vert0\rangle_{\mathbf k}\\
\label{genmfs2}
\vert u,\epsilon_F\rangle&=&\prod_{\mathbf k}\vert\Psi(\mathbf k)\rangle
\end{eqnarray}
where $\sum_a$ denotes the sum over all the $^{N_B}C_{N_F(\mathbf k)}$
combinations of the the index $a$ and $\prod_{\mathbf k}$ denotes the direct
product of the states defined at each point in the $BZ$.


Our definition of the quantum distance matrix is,
\begin{equation}
\label{ddef2}
D^2(\mathbf k_1,\mathbf k_2)\equiv 1-
\vert\langle u,\epsilon_F\vert 
E(\mathbf k_1,\mathbf k_2)
\vert u,\epsilon_F\rangle\vert^\alpha
\end{equation}
where $E(\mathbf k_1,\mathbf k_2)$ are the exchange operators. They are unitary
operators and their action of the fermion creation operators is given by,
\begin{eqnarray}
\label{eonc1}
E(\mathbf k_1,\mathbf k_2)C^\dagger_a(\mathbf k_1)
E^\dagger(\mathbf k_1,\mathbf k_2)&=&\pm C^\dagger_a(\mathbf k_2)\\
\label{eonc2}
E(\mathbf k_1,\mathbf k_2)C^\dagger_a(\mathbf k_2)
E^\dagger(\mathbf k_1,\mathbf k_2)&=&\pm C^\dagger_a(\mathbf k_1).
\end{eqnarray}
The $\pm$ signs above depend on the ordering convention of the 
creation operators in the definition of the many-body states. While it is
important to keep track of them for correlated states, as we will
see below, due to the factorized form of MFS, the distances are independent 
of the signs.

The action of the exchange operator on the states is
\begin{eqnarray}
\nonumber
E(\mathbf k_1,\mathbf k_2)\vert\psi(\mathbf k_1)\rangle\otimes
\vert\psi(\mathbf k_2)\rangle&=&\pm
\left(\sum_{aa'}\Psi_{a_1\dots a_{N_F(\mathbf k_1)}}(\mathbf k_1)\right.\\
\nonumber
&&\left.\Psi_{a'_1\dots a'_{N_F(\mathbf k_2)}}(\mathbf k_2)\right)\\
\nonumber
&&\prod_{l'=1}^{N_F(\mathbf k_2)}C^\dagger_{a_l'}(\mathbf k_2)
\vert0\rangle_{\mathbf k_2}\\
\nonumber
&&\otimes\prod_{l=1}^{N_F(\mathbf k_1)}C^\dagger_{a_l}(\mathbf k_1)
\vert0\rangle_{\mathbf k_1}.\\
\label{eonstates}
\end{eqnarray}

Equations (\ref{ddef2}), (\ref{eonc1}), (\ref{eonc2}) and (\ref{eonstates})
imply
\begin{equation}
D^2(\mathbf k_1,\mathbf k_2)=1-
\left\vert
\langle\Psi(\mathbf k_1))\vert\Psi(\mathbf k_2))\rangle
\right\vert^{2\alpha}
\delta_{N_F(\mathbf k_1),N_F(\mathbf k_2)}.
\end{equation}

Note that when $\alpha=1$, the RHS of the above equation is the Hilbert-
Schmidt distance between $\Psi(\mathbf k_1)$ and $\Psi(\mathbf k_2)$.
Thus we have shown that the quantum distances of the mean field states
reduce to the standard definition in terms of the overlap of wavefunctions.
For $\alpha=1$, it is exactly the Hilbert-Schmidt distance between the 
$N_{F(\mathbf k)}$ states at $\mathbf k$. Our definition also implies that 
the distance between two quasi-momenta with different occupation numbers is 
equal to 1.

\section{Euclidean embedding of CDW states}
\label{cdwee}

In this section we implement the procedure described in section \ref{genee}
and give an explicit solution to the problem of isometrically embedding
the distance matrix of the CDW state in a Euclidean space.

Using the definitions, the matrix of the 
squared distances, $D_{CDW}^{2}$ can be written as \cite{Paper1}:
\begin{equation}
\label{dcdwblocks}
D_{CDW}^2=\frac{3}{4}\left(\begin{array}{cc} 
{\cal I}-I&{\cal I}-I\\{\cal I}-I&{\cal I}-I\end{array}\right)
+\left(\begin{array}{cc} 0&I\\I&0\end{array}\right).
\end{equation}
Note that $ee^T={\cal I}$, where $e$ is defined below equation(\ref{gmatdef}).
The Gram matrix defined in equation(\ref{gmatdef}) is,
\begin{eqnarray}
G_{CDW}&=&-\frac{1}{2}AD_{CDW}^2A\\
A&\equiv&\left(\begin{array}{cc}I&0\\0&I\end{array}\right)
-\frac{1}{L}\left(\begin{array}{cc}{\cal I}&{\cal I}\\{\cal I}&{\cal I}
\end{array}\right).
\end{eqnarray}
It is easy to check that,
\begin{equation}
\left[A,D_{CDW}^2\right]=0~\Rightarrow~\left[A,G_{CDW}\right]=0
\end{equation}
Thus, $A$ and $G_{CDW}$ have the same eigenvectors. It is quite easy 
to construct them.  We give the answer below.

Define a complete, orthonormal set of  $\frac{L}{2}$ dimensional 
column vectors $a^{\mu}$, $\mu=1...\frac{L}{2}$, where 
$a^1=\sqrt{\frac{2}{L}}(1,1,\dots ,1)^{T}$. Further define a complete set 
of $L$ dimensional orthonormal vectors,
\begin{eqnarray}
b^{i}&\equiv& \frac{1}{\sqrt{2}} 
\left(\begin{array}{c}a^i\\a^i\end{array}\right),~~~~~~~i=1,\dots,\frac{L}{2}\\
&\equiv& \frac{1}{\sqrt{2}} 
\left(\begin{array}{c}a^{i-L/2}\\-a^{i-L/2}\end{array}\right),
~i=\frac{L}{2}+1,\dots,L.
\end{eqnarray}
It can be verified that, 
\begin{eqnarray}
G_{CDW}b^1&=&0\\
G_{CDW}b^i&=&\frac{1}{4}b^{i},~i=2,\dots,\frac{L}{2}\\
          &=&\frac{1}{2}b^{i},~i=\frac{L}{2}+1,\dots,L.
\end{eqnarray}
Thus,
\begin{equation}
\left(G_{CDW}\right)_{ij}=\sum_{k=2}^{L/2} \frac{1}{4}b^k_ib^k_j
+\sum_{k=L/2+1}^{L} \frac{1}{2}b^k_ib^k_j.
\end{equation}
The above equations and the procedure described in Section \ref{genee}
gives the explicit solution for the embedding to be the $L$, 
$(L-1)$-dimensional vectors, $x_i,~i=1,\dots,L$, with components,
$\left(x_i\right)^n,~n=1,\dots,L-1$, given by
\begin{eqnarray}
\left(x_i\right)^n&=&\frac{1}{2}b^{n+1}_i,~~~~n=1,\dots,\frac{L}{2}-1\\
                  &=&\frac{1}{\sqrt 2}b^{n+1}_i,~~n=\frac{L}{2},\dots,(L-1).
\end{eqnarray}
Thus, the distance matrix at $V=\infty$, can be isometrically embedded
in a Eulcidean space with dimension equal to $L-1$.

%

\end{document}